*Article*

# Fusion of Single and Integral Multispectral Aerial Images

**Mohamed Youssef and Oliver Bimber,***

Institute of Computer Graphics, Johannes Kepler University Linz, 4040 Linz, Austria; mohamed.youssef@jku.at
* Correspondence: oliver.bimber@jku.at; Tel.: +43-732-2468-6631

**Abstract:** An adequate fusion of the most significant salient information from multiple input channels is essential for many aerial imaging tasks. While multispectral recordings reveal features in various spectral ranges, synthetic aperture sensing makes occluded features visible. We present a first and hybrid (model- and learning-based) architecture for fusing the most significant features from conventional aerial images with the ones from integral aerial images that are the result of synthetic aperture sensing for removing occlusion. It combines the environment's spatial references with features of unoccluded targets that would normally be hidden by dense vegetation. Our method outperforms state-of-the-art two-channel and multi-channel fusion approaches visually and quantitatively in common metrics, such as mutual information, visual information fidelity, and peak signal-to-noise ratio. The proposed model does not require manually tuned parameters, can be extended to an arbitrary number and arbitrary combinations of spectral channels, and is reconfigurable for addressing different use cases. We demonstrate examples for search and rescue, wildfire detection, and wildlife observation.

**Keywords:** image fusion; aerial imaging; multispectral; synthetic aperture sensing; Airborne Optical Sectioning; occlusion removal

## 1. Introduction

Occlusion caused by dense vegetation, such as forest, represents a fundamental problem for many applications that apply aerial imaging. These include search and rescue, wildfire detection, wildlife observation, surveillance, forestry, agriculture, and archaeology. With Airborne Optical Sectioning (AOS) [1–5], we have introduced a synthetic aperture imaging technique that removes occlusion in aerial images in real time (Figure 1a–d). It computationally registers and integrates multiple (single) images captured with conventional camera optics at different drone positions into an integral image that mimics a wide (several meters) synthetic aperture camera. Thereby, the area in which the single images are sampled represents the size of the synthetic aperture. Image registration and integration depend both on the camera poses where the single images have been recorded and on a given focal surface (e.g., a defined plane or a registered digital elevation model of the ground surface). The resulting integral image have an extremely shallow depth of field. Because of this, all targets located on the focal surface appear sharp and unoccluded, while occluders not located on the focal surface appear severely defocused. In fact, the blur signal of the occluders is spread widely over the integral image, which suppresses their contribution. High spatial resolution, real-time processing capabilities, and wavelength independents are the main advantages of AOS over alternatives such as light detecting and ranging (LiDAR) or synthetic aperture radar (SAR). AOS can be applied to images captured in the visible, near-infrared, or far-infrared spectrum, and its processing is in the range of milliseconds.

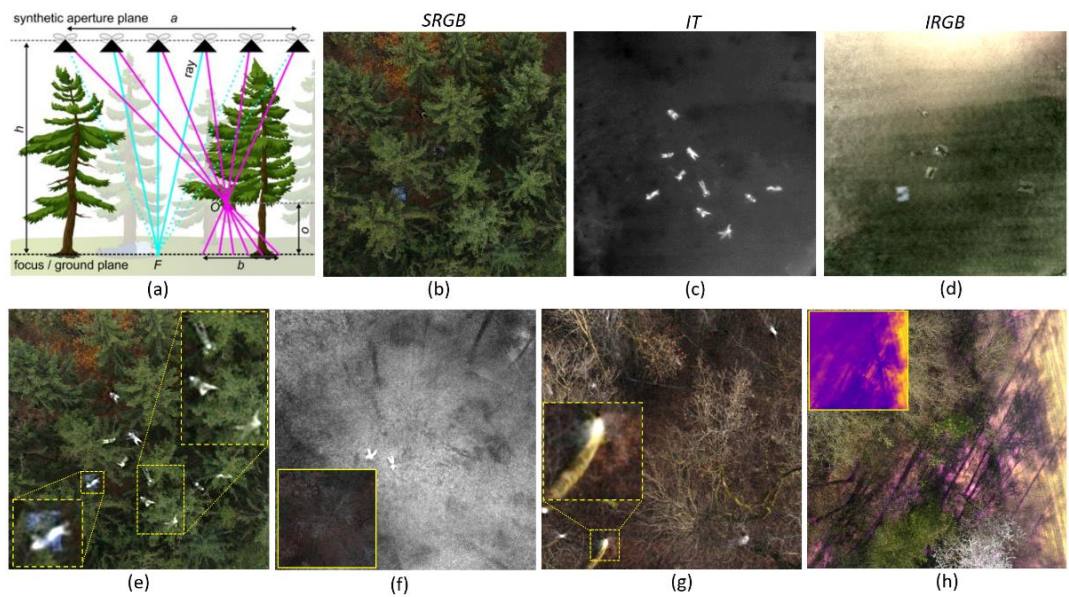

**Figure 1.** Airborne Optical Sectioning (AOS) principle (**a**): Registering and integrating multiple images captured along a synthetic aperture of size a while computationally focusing on focal plane F at distance h will defocus occluders O at distance o from F (with a point-spread of b) while focusing targets on F. Single RGB image, SRGB (**b**); integral thermal image, IT (**c**); and integral RGB image, IRGB (**d**) of dense forest captured with a square synthetic aperture area (30 m × 30 m) with 1 m × 3 m dense sampling while computationally focusing on the forest ground. All images show the same scene, at the same time, from the same pose. (**e**) Fused result from (**b**–**d**). Close-ups are in dashed boxes. (**f**) Fusion of IT and IRGB under dark light conditions—increasing exposure through integration. Original SRGB image in solid box. (**g**) SRGB and IT fusion of sparse forest—highlighting the bottoms of tree trunks in the thermal channel. Close-up in dashed box. (**h**) SRGB and color-coded IT fusion—revealing hot ground patches in the IT channel. Color-coded IT image in sloid boxes.

One major limitation of these integral images is their lack of spatial references, as the surrounding environment's details (e.g., features of the forest structure, such as distinct trees, sparse and dense regions, etc.) are suppressed in the shallow depth of field. Such spatial references, however, are important for orientation while inspecting the images. In this article, we propose a novel hybrid (model- and learning-based) architecture for fusing the most significant (i.e., the most salient) information from single and integral aerial images into one composite image (Figure 1e–h). This composite image combines the environment's spatial references provided by single images (for orientation; not visible in integral images) with features of the unoccluded targets provided by integral images (not visible in single aerial images). Our method is the first to fuse visible and occluded features in multiple spectral bands, outperforms the state of the art (visually and quantitatively), does not require manually tuned parameters, can be extended to an arbitrary number and arbitrary combinations of spectral channels, and is reconfigurable to address different use cases.

Classically, the fusion of lower-resolution hyperspectral and higher-resolution multispectral images is concerned with up-sampling, sharpening, and super-resolution [6]. New image fusion approaches that extract and combine salient image features, especially for recordings in the infrared and visible spectral ranges, have become an active research field in recent years. Feature extraction and the fusion strategy itself are two main components of such methods. Multi-scale transform-based approaches [7–10] split images into multiple scales for the analysis of both fine and coarse features. Low-rank-based methods [11,12] decompose the image matrix into low-rank and sparse representations. Such model-based approaches often require a lot of computational resources and extensive manual adjustments. In [13], a hybrid concept is followed, which first extracts features using a model-based approach and then feeds these features as input to a learning-based approach (i.e., a pretrained neural network). Usually, convolutional neural networks (CNNs) or generative adversarial networks (GANs) are architectures of choice for image fusion today. CNNs are used to extract local and global features and generate the fused image [14–18]. However, despite their remarkable fusion performance over model-based approaches, deeper networks often lose important details due to stacking a series of pooled convolution layers. Transformer architectures [19–22] can overcome this by extracting complementary features from the input images. With GANs, the problem of a limited number of ground-truth data for infrared-visible image fusion can be overcome. The GAN learning scheme [23] depends on the generator and discriminator, and such architectures discriminate between the generated fused image and the input images [24]. However, using only one discriminator may lead to networks being biased towards favoring either the generated fused images or the input images. Therefore, dual discriminators were adopted [25,26] to ensure high-quality fusion results.

The main contributions of this work are as follows:

(1) We present the first fusion approach for multispectral aerial images that combines the most salient features from conventional aerial images and integral images which result from synthetic aperture sensing. While the first contains the environment's spatial references for orientation, the latter contains features of unoccluded targets that would normally be hidden by dense vegetation. Our model does not require manually tuned parameters, can be extended to an arbitrary number and arbitrary combinations of spectral channels, and is reconfigurable for addressing different use cases. This method is explained in Section 2.

(2) Our method outperforms state-of-the-art two-channel and multi-channel fusion approaches visually and quantitatively in common metrics, such as mutual information, visual information fidelity, and peak signal-to-noise ratio. We demonstrate results for various use cases, such as search and rescue, wildfire detection, and wildlife observation. These results are presented in Section 3.

## 2. Materials and Methods

Our proposed architecture is depicted in Figure 2. It is motivated by the hybrid approach in [13], but instead of relying on simple image averaging for the final fusion process (which does not lead to better results than alpha blending, as shown in Figure 4), we integrate feature-masked input channels. Our architecture has $f + 1$ input channels, whereby one channel takes a basis image ($B$) that remains unmodified to provide spatial references. The other channels take images ($F_1 ... F_f$) from which features need to be extracted and fused with the basis image. Depending on the use case, the input to the channels can vary. For instance, single RGB (SRGB) images were used for the basis channel in Figure 1e,g,h, while integral thermal (IT) images were used instead in Figure 1f; raw integral images were used in the feature channels in Figure 1e–g, while color-coded IT was used instead in Figure 1h. Integral RGB (IRGB) and SRGB images were used in the feature channels. Note that the fusion process is the same for more than two feature channels.

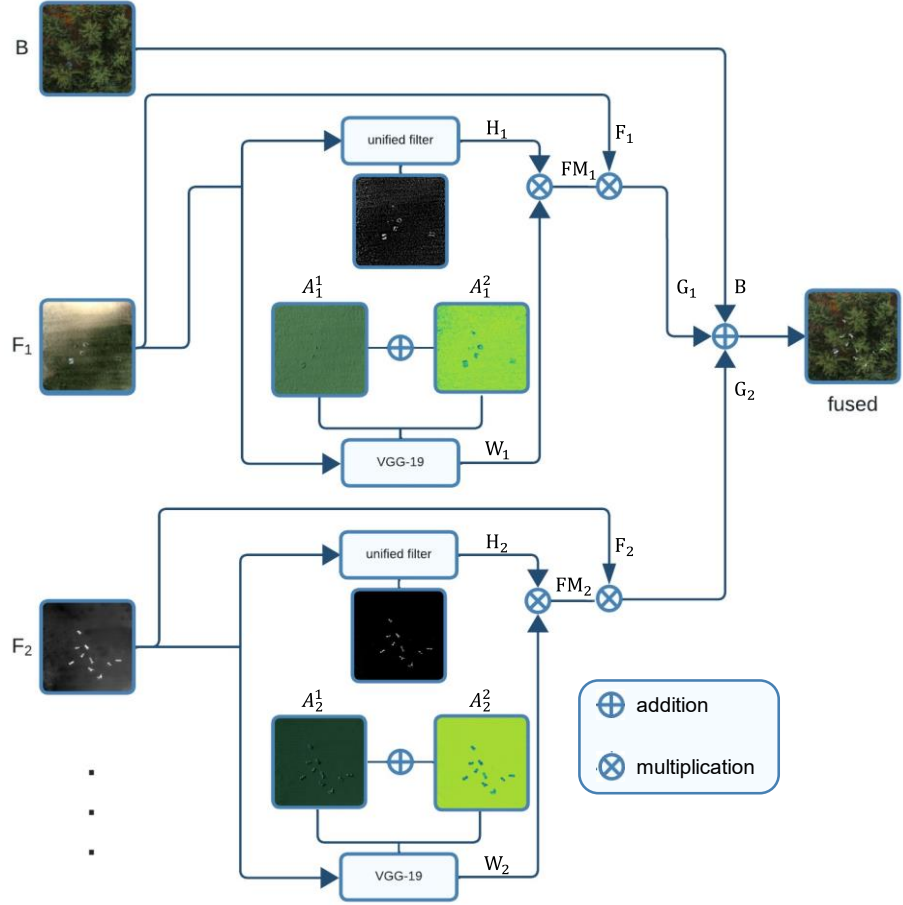

**Figure 2.** Proposed hybrid fusion architecture. Multiple input channels (one basis channel that remains unmodified and provides spatial references, and an arbitrary number of additional channels from which salient features are extracted) are fused into one composite image. Each feature channel applies multiple model-based and learning-based feature extractors (unified filters and VGG-layers in our case). Variables refer to Equations (1)–(7).

As shown in Figure 2, each $F_n$-branch splits into two parallel branches for feature extraction: one with a unified filter [27] for model-based feature extraction and one with the pretrained very deep convolutional network, VGG-19 [28], for learning-based feature extraction.

The unified filter determines the high-detail part ($H_n$) of the input channel ($F_n$) as follows:

$$H_n = F_n - arg\ min_{H_n}\ \|F_n - H_n\|_2^2 + \lambda\ (\|g_x * H_n\|_2^2 + \|g_y * H_n\|_2^2), \quad (1)$$

where $n \epsilon \{1, 2, \ldots, f\}$; we always used the default of $\lambda = 5$, as optimized in [12], and $g_x = [-1\ \ 1]$ and $g_y = [-1\ \ 1]^T$ are the horizontal and vertical convolution gradient operators, respectively.

The pretrained VGG-19 network architecture shows remarkable results in feature extraction, which is usually applied for classification tasks. Consider that $\emptyset_n^i$ indicates the features map extracted from $n$-th channel in the $i$-th VGG-19 layer:

$$\emptyset_n^i = \Phi_i(F_n), \quad (2)$$

where $\Phi_i(.)$ denotes an $i$-th layer. Deeper layers in VGG detect high-level features, while lower-level layers detect abstract features such as edges and colors. We decided to use only the first two layers (i.e., $i \epsilon \{1, 2\}$—which represents *relu_1_1* and *relu_2_1*, respectively), as we are more interested in the abstract, low-level salient features from $F_n$ rather than its high-level features.

After determining the features $\emptyset_n^i$, an activity level map $A_n^i$ is calculated by the $l_1-norm$:

$$A_n^i(x,y) = \sum_x \sum_y \sum_c |\emptyset_{n\ x,y,c}^i|, \quad i \in \{1, 2\}. \tag{3}$$

where $(x, y)$ is the position in the feature map and $c$ is the channel number.

Afterwards, a weighted average map ($W_n$) is calculated by a soft-max operator:

$$W_n(x,y) = \sum_{l=1}^{k} \frac{A_n^i(x,y)}{\sum_{j=1}^{k} A_n^j(x,y)}, \quad i \in \{1, 2\}, \tag{4}$$

where $k$ denotes the number of the activity level map.

The initial feature mask ($FM_n$) is calculated by multiplying it with $W_n$ as follows:

$$FM_n(x,y) = \sum_{n=1}^{f} W_n(x,y) \cdot H_n(x,y), \tag{5}$$

and the final feature map ($G_n$) is then:

$$G_n(x,y) = FM_n(x,y) \cdot F_n(x,y), \tag{6}$$

where $F_n$ are the input feature channels, as described above (see Figure 2).

Finally, the feature maps from all the feature channels are fused with the basis channel:

$$fused(x,y) = B(x,y) + \sum_{n=1}^{f} G_n(x,y). \tag{7}$$

Our goal with this architecture is to preserve the salient information from all input channels. This, however, can only be achieved if multiple feature extractors (unified filters and VGG in our case) are combined. Applying them independently fails to remove background noise and enhance the essential target features, as demonstrated in Figure 3.

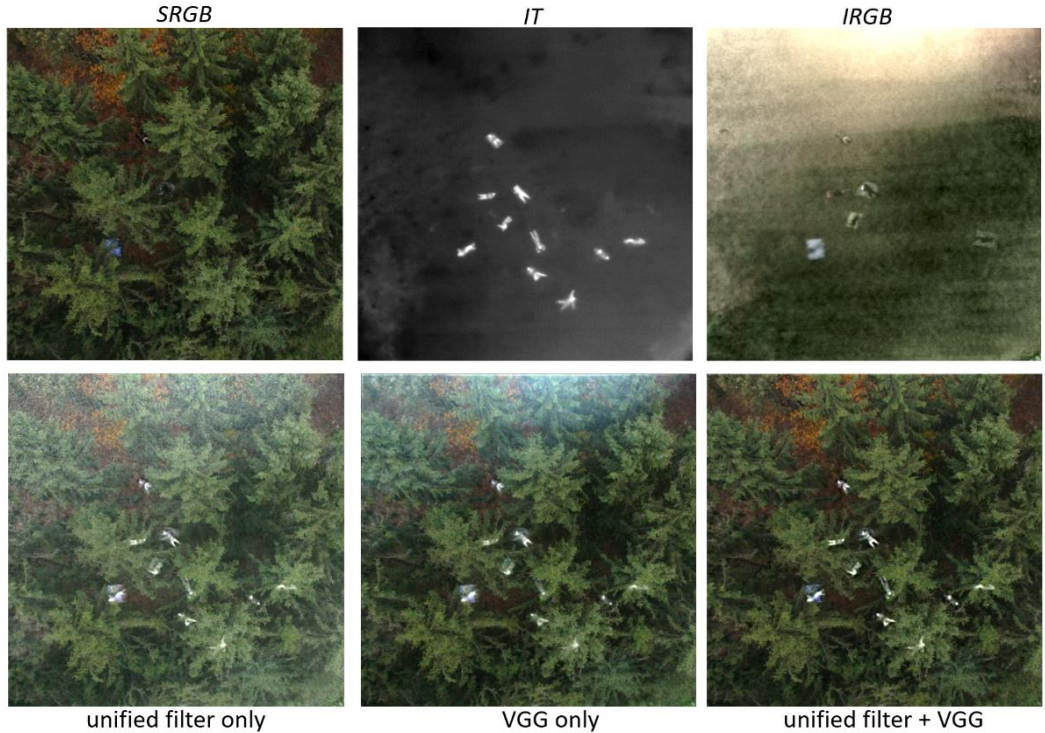

**Figure 3.** Comparison of fusion results with unified filter only, VGG only, and the combination of unified filter and VGG. Only in the latter case background noise and sampling artifacts can be removed efficiently and essential target features are enhanced. The input images (*SRGB*, *IT*, *IRGB*) show the same scene, at the same time, from the same pose.

On a 3.2 GHz Intel Core CPU with 24 GB RAM and GPU GTX 1060 6 GB, our python implementation of the proposed fusion architecture requires (for 512 × 512 px input images), approx. 1.2 s per feature channel plus 10 ms for fusing the feature channel with the basis channel. See supplementary material section for code and sample images.

## 3. Results and Discussion

Figure 4 presents a visual comparison of our approach with several state-of-the-art image fusion techniques for various scenes in a search-and-rescue use case (with data from [3]): (c) is an open field without occlusion, (a,b,d,e,f) are forest patches with different types of vegetation and densities, (a,b,c,f) show people lying on the ground, and (d,e) are empty forests. Useful features to be fused are contained in SRGB, IT, and IRGB for (a,b,f), in IT and IRGB for (c), and in SRGB and IT for (d,e). The IT image is color-coded in (e,f). For all integral images (IT and IRGB) the synthetic focal plane was set to the ground surface.

Figure 5 illustrates a visual comparison of our approach with the state of the art for wildlife observation (nesting observation with data from [2]) and wildfire detection and monitoring (with data from [29]). Fused are IRGB and color-coded IT for nesting locations with breeding birds at lower layers just below tree crowns (a), and SRGB of tree vegetation and color-coded IT of ground fires (b,c).

While some of the evaluated fusion methods support only two input channels [16,19–21], the work in [13] scales to multiple channels. In cases only two channels are supported but three channels could be used, we selected SRGB and IT. In cases where only grayscale input is supported [19,20] but color input is provided, we convert color images to grayscale.

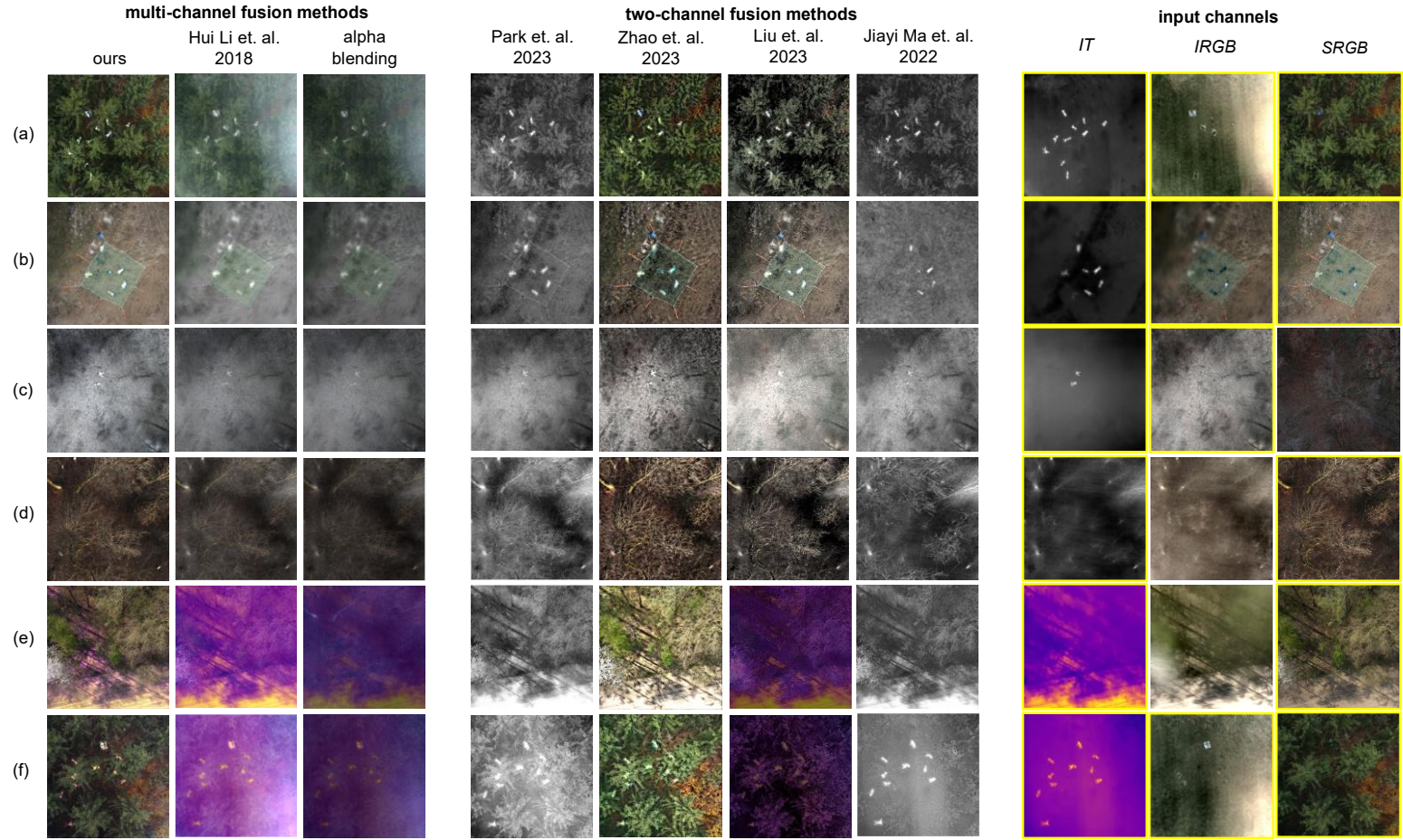

**Figure 4.** Comparison of our approach with several state-of-the-art two- and multi-channel image fusion techniques for various scenes in a search-and-rescue use case. Input channels (*SRGB*, *IT*, *IRGB*) indicated with a yellow solid box contain useful features to be fused. They show the same scene, at the same time, from the same pose. The data used for our experiments and details on how they were recorded can be found in [3]. All images, except for (**c**), have been brightness-increased by 25% for better visibility. References: Hui Li et. al. 2018 [13], Park et. al. 2023 [19], Zhao et. al. 2023 [16], Liu et. al. 2023 [21], and Jiayi Ma et. al. 2022 [20].

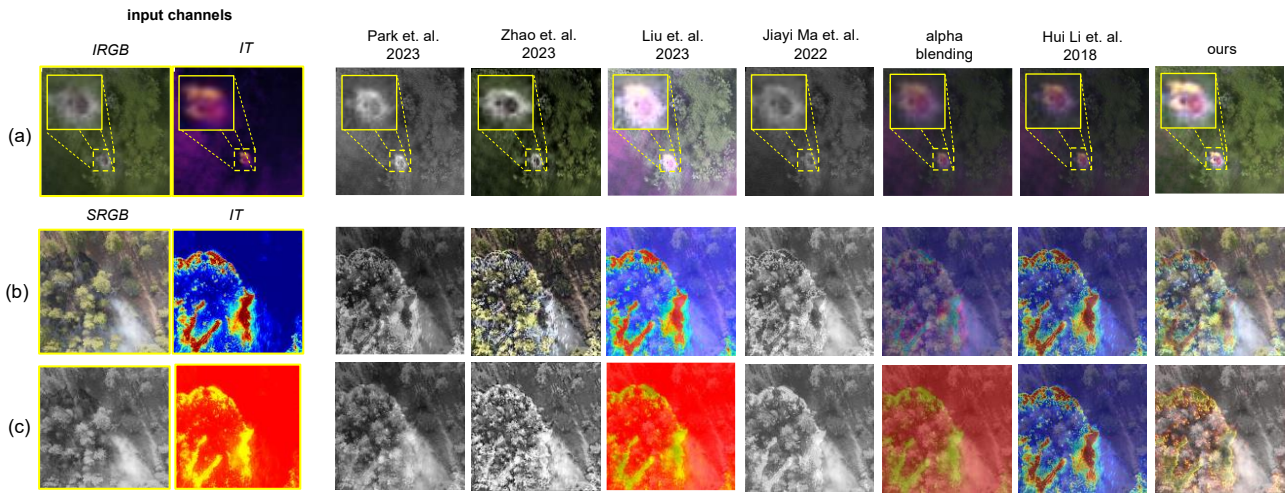

**Figure 5.** Comparison of our approach with the state of the-art for wildlife observation (**a**) and wildfire detection and monitoring (**b**,**c**) use cases. Input channels (*SRGB*, *IT*, *IRGB*) are indicated with a yellow solid box. The example in (**a**) shows results from nesting observations of breeding herons. The data are taken from [2]. The data used for the examples in (**b**,**c**) are taken from the FLAME 2 dataset [29], while structure-from-motion and multi-view stereo [30] was applied for pose estimation of each video frame. Integral thermal images (IT) are computed from single thermal video frames with [1]. The IT channel is color-coded in all examples. References: Park et. al. 2023 [19], Zhao et. al. 2023 [16], Liu et. al. 2023 [21], Jiayi Ma et. al. 2022 [20], and Hui Li et. al. 2018 [13].

The poor quality of simple image processing, such as alpha blending with an equal contribution per input channel, clearly calls for more sophisticated approaches that are capable of extracting and fusing the essential image features. The only image fusion method that is scalable to multiple (more than two) input channels [13] does not provide significantly better results than alpha blending in our experiments. The reason for this is that, in spite of using VGG-19 for feature extraction, the final fusion process relies on image averaging.

All newer methods support only two input channels and still fail to differentiate between essential image features and other high-gradient features that are considered unimportant (e.g., noise and sampling artifacts). Consequently, the fused results appear unnatural and do not reflect properly on the salient image features and details of the original input channels. Although [20] uses transformers for global feature extraction, it fails to extract the local features correctly due to the lack of sufficient correlation between neighboring pixels. In [16], a multi-model was trained for fusion and detection tasks that mutually shared learned features between the two models (i.e., fusion and detection). It struggles in generalizing to our fusion problem and fails to preserve salient information in our data as it is trained on a very different benchmark dataset [31] that is specialized to infrared and visible fusion for object detection. The work in [19] cannot accurately estimate complementary information between input images if target objects are discriminative in both images. The architecture in [21] fails in generating the saliency mask, as it was only trained on a small dataset [32] for extracting the mask, which leads to a poor fusion output at the end.

For a quantitative comparison of our results with results of the state of the art, we apply three metrics that are commonly used for evaluating image fusion techniques:

1. *Mutual Information (MI)* [33] measures the amount of information transmitted from source images (i.e., the basis and feature channels in our case) to the fused result by the Kullback–Leibler divergence [34].
2. *Visual Information Fidelity (VIF)* [35] assesses the quality of the fused result based on its fidelity, which determines the amount of information obtained from the source images.

3. *Peak Signal-to-Noise Ratio (PSNR)* [36] quantifies the ratio of peak power to the noise power in the fused result.

As shown in Table 1, our model is always superior compared to state-of-the-art two-channel fusion techniques.

**Table 1.** Quantitative comparison with two-channel fusion techniques. The best results are bolded, and the second-best results are underlined. Labels refer to corresponding scenes presented in Figures 4 and 5, respectively. **AVG** is the average over all scenes. Higher values indicate higher quality of the fused image with respect to the corresponding metric.

| | (Figure 4a) | (Figure 4b) | (Figure 4c) | (Figure 4d) | (Figure 4e) | (Figure 4f) | (Figure 5a) | (Figure 5b) | (Figure 5c) | AVG |
|---|---|---|---|---|---|---|---|---|---|---|
| **VIF** | | | | | | | | | | |
| Park et al. [19] | 0.936 | 0.675 | 0.879 | 0.667 | 0.718 | 0.940 | 0.998 | 0.452 | 0.600 | 0.762 |
| Zhao et al. [16] | 0.849 | 0.563 | 0.628 | 0.638 | 0.550 | 0.899 | 0.849 | 0.428 | 0.473 | 0.653 |
| Liu et al. [21] | 0.892 | 0.701 | 0.698 | 0.689 | 0.530 | 0.770 | 1.015 | 0.427 | 0.235 | 0.662 |
| Jiayi Ma et al. [20] | 0.898 | 0.129 | 0.912 | 0.175 | 0.692 | 0.624 | 0.884 | 0.539 | 0.119 | 0.552 |
| ours | **1.069** | **1.002** | **1.024** | **0.774** | **0.849** | **1.072** | **1.047** | **0.635** | **0.742** | **0.912** |
| **MI** | | | | | | | | | | |
| Park et al. [19] | 0.547 | 0.505 | 1.266 | 0.831 | 1.016 | 0.420 | 0.525 | 0.739 | 0.937 | 0.754 |
| Zhao et al. [16] | 0.796 | 0.388 | 1.119 | 0.708 | 1.161 | 0.901 | 0.514 | 0.584 | 0.637 | 0.756 |
| Liu et al. [21] | 0.643 | 0.552 | 1.444 | 0.606 | 1.194 | 0.634 | 0.605 | 1.023 | 0.750 | 0.828 |
| Jiayi Ma et al. [20] | 0.597 | 0.063 | 1.832 | 0.424 | 1.022 | 0.422 | 0.846 | 0.935 | 0.371 | 0.724 |
| ours | **1.207** | **1.444** | **1.843** | **1.121** | **1.346** | **1.094** | **1.526** | **1.210** | **1.281** | **1.341** |
| **PSNR** | | | | | | | | | | |
| Park et al. [19] | 20.942 | 20.509 | 17.915 | 19.475 | 13.723 | 12.849 | 20.336 | 11.250 | 12.455 | 16.606 |
| Zhao et al. [16] | 24.642 | 19.808 | 16.912 | 21.904 | 15.049 | 16.819 | 19.792 | 10.728 | 10.921 | 17.397 |
| Liu et al. [21] | 20.923 | 18.110 | 12.495 | 21.100 | 14.298 | 16.004 | 19.258 | 12.111 | 7.084 | 15.709 |
| Jiayi Ma et al. [20] | 24.791 | 18.304 | 20.792 | 20.989 | 14.460 | 14.051 | 19.535 | 12.003 | 10.791 | 17.301 |
| ours | **26.503** | **23.523** | **21.899** | **26.850** | **16.514** | **18.865** | **25.501** | **12.917** | **12.861** | **20.603** |

In contrast to two-channel fusion techniques, our method is scalable with respect to the number of input channels to be fused. The comparison with multi-channel techniques in Table 2 also reveals its clear advantage over the state of the art. See supplementary material section for results.

**Table 2.** Quantitative comparison with multi-channel fusion techniques. The best results are bolded, and the second-best results are underlined. Labels refer to corresponding scenes presented in Figures 4 and 5, respectively. **AVG** is the average over all scenes. Higher values indicate higher quality of the fused image with respect to the corresponding metric.

| | (Figure 4a) | (Figure 4b) | (Figure 4c) | (Figure 4d) | (Figure 4e) | (Figure 4f) | (Figure 5a) | (Figure 5b) | (Figure 5c) | AVG |
|---|---|---|---|---|---|---|---|---|---|---|
| **VIF** | | | | | | | | | | |
| alpha blending | 0.623 | 0.685 | 0.779 | 0.579 | 0.447 | 0.422 | 0.616 | 0.434 | 0.524 | 0.568 |
| Hui Li et al. [13] | 0.732 | 0.797 | 0.803 | 0.627 | 0.515 | 0.434 | 0.637 | 0.462 | 0.562 | 0.619 |
| ours | **1.068** | **1.040** | **1.023** | **0.774** | **0.863** | **1.088** | **1.047** | **0.635** | **0.742** | **0.920** |
| **MI** | | | | | | | | | | |
| alpha blending | 1.006 | 1.230 | 1.231 | 0.909 | 0.877 | 1.134 | 0.699 | 0.828 | 1.245 | 1.018 |
| Hui Li et al. [13] | 1.020 | 1.245 | 1.228 | 0.904 | 1.190 | 1.102 | 0.675 | 0.939 | 1.189 | 1.055 |
| ours | **1.149** | **1.940** | **1.837** | **1.121** | **1.419** | **1.173** | **1.526** | **1.210** | **1.281** | **1.406** |
| **PSNR** | | | | | | | | | | |
| alpha blending | 18.827 | 21.461 | 18.138 | 24.832 | 15.227 | 14.938 | 20.080 | 11.032 | 11.068 | 17.289 |
| Hui Li et al. [13] | 16.307 | 19.291 | 18.131 | 24.800 | 14.130 | 14.930 | 19.987 | 9.511 | 9.112 | 16.244 |
| ours | **20.581** | **22.068** | **21.350** | **26.850** | **16.695** | **16.904** | **25.501** | **12.917** | **12.861** | **19.525** |

## 4. Conclusions

An adequate fusion of the most significant salient information from multiple input channels is essential for many aerial imaging tasks. While multispectral recordings reveal features in various spectral ranges, synthetic aperture sensing makes occluded features visible. Our proposed method effectively fuses all these features into one composite image which provides important spatial reference queues for orientation and reveals hidden target objects in addition. It outperforms state-of-the-art image fusion approaches visually and quantitatively, is extendable to an arbitrary number of input channels, is easy to use as it does not require manually tuned parameters, and is reconfigurable to address different use cases. Such use cases include search and rescue, soil moisture analysis, wildfire detection and monitoring, observation and tracking of wildlife, surveillance, border control, and others. Occlusion caused by dense vegetation is often a limiting factor for these tasks.

In addition to VGG-19, we evaluated ResNet-50 [37] for feature extraction. However, it turned out that the lower VGG-19 layers extract features significantly better than ResNet-50. This might also be a reason for VGG-19 outperforming ResNet-50 in other domains, such as segmentation and classification in medical images [38]. In future, we want to explore whether extending our architecture by more than two feature-extractor branches per feature channel further improves results and how to achieve real-time performance.

**Supplementary Materials:** The source code and sample images can be downloaded at: https://doi.org/10.5281/zenodo.10450971. Additional information is available at: https://github.com/JKU-ICG/AOS/, (accessed on 5 February 2024).

**Author Contributions:** Conceptualization, O.B.; methodology, O.B. and M.Y.; software, M.Y.; validation, O.B. and M.Y.; formal analysis, O.B. and M.Y.; investigation, O.B. and M.Y.; resources, M.Y.; data curation, O.B. and M.Y.; writing—original draft preparation, O.B. and M.Y.; writing—review and editing, O.B. and M.Y.; visualization, O.B. and M.Y.; supervision, O.B.; project administration, O.B.; funding acquisition, O.B. All authors have read and agreed to the published version of the manuscript.

**Funding:** This research was funded by the Austrian Science Fund (FWF) and the German Research Foundation (DFG) under grant numbers P 32185-NBL and https://doi.org/10.55776/I6046, as well as by the State of Upper Austria and the Austrian Federal Ministry of Education, Science and Research via the LIT-Linz Institute of Technology under grant number LIT2019-8-SEE114. Open Access Funding by the University of Linz.

**Data Availability Statement:** The data presented in this study are openly available in Zenodo at https://doi.org/10.5281/zenodo.10450971, reference number [10450971], (accessed on 5 February 2024).

**Conflicts of Interest:** The authors declare no conflicts of interest.